# Optically and magnetically addressable valley pseudospin of interlayer excitons in bilayer MoS$_2$

Yanchong Zhao[1,2], Luojun Du[3,*], Shiqi Yang[4,5], Jinpeng Tian[1,2], Xiaomei Li[1,2], Cheng Shen[1,2], Jian Tang[1,2], Yanbang Chu[1,2], Kenji Watanabe[6], Takashi Taniguchi[7], Rong Yang[1,8,9], Dongxia Shi[1,2,8], Zhipei Sun[3,10], Yu Ye[4,*], Wei Yang[1,2,9,*], Guangyu Zhang[1,2,8,9,*]

[1]Beijing National Laboratory for Condensed Matter Physics; Key Laboratory for Nanoscale Physics and Devices, Institute of Physics, Chinese Academy of Sciences, Beijing, 100190, China

[2]School of Physical Sciences, University of Chinese Academy of Science, Beijing 100190, China

[3]Department of Electronics and Nanoengineering, Aalto University, Tietotie 3, FI-02150, Finland

[4]State Key Laboratory for Mesoscopic Physics and Frontiers Science Center for Nano-optoelectronics, School of Physics, Peking University, Beijing 100871, China

[5]School of Physics and Academy for Advanced Interdisciplinary Studies, Peking University, Beijing 100871, China

[6]Research Center for Functional Materials, National Institute for Materials Science, 1-1 Namiki, Tsukuba 305-0044, Japan

[7]International Center for Materials Nanoarchitectonics, National Institute for Materials Science, 1-1 Namiki, Tsukuba 305-0044, Japan

[8]Beijing Key Laboratory for Nanomaterials and Nanodevices, Beijing 100190, China

[9]Songshan-Lake Materials Laboratory, Dongguan, Guangdong Province 523808, China

[10]QTF Centre of Excellence, Department of Applied Physics, Aalto University, FI-00076 Aalto, Finland

*Corresponding authors. Email: luojun.du@aalto.fi; ye_yu@pku.edu.cn; wei.yang@iphy.ac.cn; gyzhang@iphy.ac.cn

**Interlayer valley excitons in bilayer MoS$_2$ feature concurrently large oscillator strength and long lifetime, and hence represent an advantageous scenario for valleytronic applications. However, control of valley pseudospin of interlayer excitons in pristine bilayer MoS$_2$, which lies at the heart of valleytronics, has remained elusive. Here we report the observation of highly circularly polarized photoluminescence from interlayer excitons of bilayer MoS$_2$ with both optical and magnetic addressability. Under excitation of circularly polarized light near exciton resonance, interlayer excitons of bilayer MoS$_2$ show a near-unity, but negative circular polarization. Significantly, by breaking time-reversal symmetry with an out-of-plane magnetic field, a record level of spontaneous valley polarization (7.7%/Tesla) is identified for interlayer excitons in bilayer MoS$_2$. The giant valley polarization of the interlayer excitons in bilayer MoS$_2$, together with the feasibility of electrical/optical/magnetic control and strong oscillator strength, provides a firm basis for the development of next-generation electronic and optoelectronic applications.**

The control of valley degree of freedom, which labels the local degenerate energy extrema in momentum space, opens up a new paradigm to encode and process binary information for exotic valleytronics, an appealing alternative to the conventional charge-based electronics and spintronics[1-4]. Monolayer transition metal dichalcogenides (TMDCs) are promising candidates for valley-functional electronics and optoelectronics, because of the optically addressable valley physics and large exciton oscillator strength[4-6]. However, owing to the strong intervalley electron-hole exchange interactions, the valley lifetime of exciton in monolayer TMDCs is very short, only a few picoseconds even at low temperatures[7-11]. Such ultrafast valley depolarization imposes grand challenges to the valley-based information storage and processing applications.

Thanks to the spatial separation between the wave functions of electrons and holes, interlayer excitons (IXs) have a rather weak exchange interaction and thus provide an attractive opportunity to circumvent the limitations of short valley lifetime and fast intervalley relaxation. By engineering the type-II band alignment[12], IXs with long valley lifetimes and robust valley polarization have been demonstrated across the atomically sharp interfaces in a wide variety of TMDC heterostructures[4,11,13-16]. Although the IXs of TMDC heterostructures have largely advanced the valleytronics[4,13,17-23], the rather weak optical absorption of IXs limits further developments[4,24].

Apart from TMDC heterostructures, recent progresses uncover that 2H Mo-based TMDC bilayers or multilayers can also feature IXs[24-32]. In the following, we omit "2H" for simplicity. Remarkably, IXs in Mo-based bilayer TMDCs can harbour the merits of both excitons in monolayer TMDCs (e.g., large oscillator strength) and IXs in TMDC heterostructures (e.g., long valley lifetime and large out-of-plane electric dipole moment), showing crucial advantages for realizing novel valleytronic devices[24,33]. However, current researches are mainly focused on revealing the existence of IXs in bilayer TMDCs through absorption spectra or photoluminescence (PL) emission under off-resonance excitation[24,29-31]. The control of valley pseudospin of IXs in bilayer TMDCs and the further realization of large valley polarization with optical/magnetic addressability, which are the cornerstones of valleytronics, have not yet been unveiled.

In this work, we report the optical and magnetic addressability of valley pseudospin for IXs in bilayer $MoS_2$. Under excitation of circularly polarized light near exciton resonance, we demonstrate that IXs in bilayer $MoS_2$ show a close-to-unity circular polarization, but with chirality opposite to the excitation photons. In addition, by breaking the valley degeneracy with an out-of-plane magnetic field, we observe a record level of spontaneous valley polarization (~ 7.7%/Tesla) for IXs in bilayer $MoS_2$. The giant valley polarization of IXs in bilayer $MoS_2$ with both optical and magnetic addressability, together with the long valley lifetime, strong oscillator strength and large out-of-plane electric dipole, provides a firm basis for the development of valley-functional electronic and optoelectronic applications.

To demystify the valley-contrasting properties for IXs in bilayer $MoS_2$, we fabricate high-quality hexagonal boron nitride (*h*-BN) encapsulated devices with two symmetric few-layer graphene gates, as illustrated in Fig. 1A. Three *h*-BN encapsulated bilayer $MoS_2$ devices (labeled as D1, D2 and D3) are studied. All three devices show the same behavior. Unless otherwise specified, the data presented here are taken from device D1 in a high vacuum at 10 K. For device D1, the thicknesses of top and bottom *h*-BN gate dielectrics are same and ~ 9.5 nm (fig. S1). By applying the same (opposite) voltages on the two graphene gates, it varies the carrier density (vertical electric field) in bilayer $MoS_2$ while keeping the electric field (doping) unchanged. Figure 1C presents the color plot of PL spectra as a function of out-of-plane electric field $E_z$ under 2.33 eV off-resonance excitation. Prominent IX emissions can be identified, besides the well-known intralayer A exciton transition around 1.92 eV. At zero electric field, IXs are about 70 meV above the intralayer A excitons, in good line with previous results[24,30]. It is noteworthy that the intensity of IXs at $E_z$ =0 is about 16% of that of the intralayer A excitons, which shows an enhancement of more than 60% compared to the recent results[30]. The observed large emission intensity of IXs may benefit from the high quality of our samples.

On applying an out-of-plane electric field, the IXs are split into two well-separated branches, labeled as IX1 and IX2, respectively. Note that a positive $E_z$ is defined as the situation of positive/negative top/back gate voltage, and vice versa. The transition energies of IX1 (IX2) increase (decrease) linearly with the external electric fields, evidencing the out-of-plane static electric dipole and quantum-confined Stark effect. Through linear fitting (black dotted lines in Fig. 1C), we extract the electric dipole moments of $\mu = (0.56 \pm 0.01)\ e\ nm$ for IX1 and IX2, which are consistent with recent results[24,30] and the calculated value ($0.605\ e\ nm$) based on hybridized hole wavefunctions (Supplementary Section 2). Note that the energies of intralayer A excitons remain barely unchanged with an applied electric field, indicating the vanishing electric dipole and in close analogy to the excitons in monolayers[34]. The splitting of IXs can be understood that there are two IX species in bilayer MoS$_2$ (Fig. 1B): one with electron localized in lower layer (IX1) and the other with electron resided in upper layer (IX2). Due to the layer degeneracy of band structure, IX1 and IX2 are energy degenerated under zero electric field. When applying an external electric field $E_z$, it introduces an asymmetric interlayer electrostatic potential $\Phi = -edE_z$ between the upper and lower layers (*e* and *d* denote the elementary charge and interlayer distance, respectively), breaking the layer degeneracy and thus causing a global first-order Stark shift of energy band[24,35,36]. Consequently, the two energetically degenerate IX1 and IX2 split in energy and show giant Stark effect. The energy splitting between IX1 and IX2 can be tuned over 100 meV, in fair agreement with the recent results[24,30].

To reveal the valley-contrasting properties of IXs in bilayer MoS$_2$, we tune the excitation photon energy to 1.96 eV (633 nm) near exciton resonance. Figure 1D shows the color plot of non-polarized PL spectra as a function of electric field $E_z$, excited by 1.96 eV on-resonance excitation. Being akin to the case under 2.33 eV off-resonance excitation (Fig. 1C), IXs splitting into two well-separated branches at finite electric field (labeled as IX3 and IX4) can also be unequivocally observed under 1.96 eV on-resonance excitation (Fig. 1D). IX3 and IX4 belong to the same type IX, but with electrons localized in the opposite layers (Fig. 1B). Via linear fitting (black dotted lines in Fig. 1D), the electric dipole moments of IX3 and IX4 are extracted, e.g., $\mu = (0.55 \pm 0.01)\ e\ nm$, which are in good line with that of IX1/IX2. Figure 1F plots the transition energies against the applied electrical field for IX2 and IX4 extracted from Figs. 1C and 1D, respectively. Remarkably, the transition energies of IX4 are 40 meV smaller than that of IX2. While the energies of intralayer A excitons under 1.96 eV excitation are same with that under 2.33 eV excitation (Fig. 1E). This indicates that IXs measured under 1.96 eV on-resonance excitation are new species and different from those observed under 2.33 eV excitation. Interestingly, the emission intensities of such new IX species under 1.96 eV on-resonance excitation are much stronger than that of IXs observed under 2.33 eV excitation and can even be larger than that of intralayer A excitons (Fig. 1E). Moreover, the emission intensities of IXs under 1.96 eV on-resonance excitation are highly tunable (Fig. 1G). When the energies of IXs are tuned to the vicinity of intralayer A excitons or trions[37], the intensities of IXs are strongly enhanced. It is noteworthy that trions have been interpreted as Fermi polaron-polaritons in some studies[38]. Such highly unusual enhancement of emission intensities for IXs under 1.96 eV excitation may origin from the resonance energy transfer from intralayer A transitions or trions/Fermi polarons and deserves further in-depth studies to reveal the detailed microscopic mechanism[39].

Owing to the out-of-plane mirror symmetry breaking of bilayer MoS$_2$, both spin-singlet and spin-triplet IXs are optically bright in principle[21,22,40]. For the two different IXs observed under 1.96 eV and 2.33 eV excitations, is it possible that one is a spin-singlet species, while the other is a spin-triplet exciton? To verify this hypothesis, we perform the measurements of valley Zeeman effect. Valley Zeeman splitting provides the key characteristics of valley/spin/intra-atomic orbital angular momentum and can effectively distinguish IXs between spin-singlet and spin-triplet species[20-22,41-44]. Figure 2A sketches the Zeeman shift of the band edges, including contributions from the spin, valley orbital and intra-atomic *d* orbital magnetic moments. For clarity, we show the shifts exaggerated. Please refer to the details in Supplementary Section 6 for the valley-dependent polarization selection rules of the intralayer A excitons, spin-singlet and spin-triplet IXs in bilayer MoS$_2$. It is noteworthy that because of the strong spin-layer locking effects, the impact of interlayer hybridization on valley Zeeman splitting in bilayer MoS$_2$ is marginal and thus can be ignored[35,45-48]. For intralayer A excitons, only the spin-singlet species is bright and the corresponding valley Zeeman splitting under an applied magnetic field $B$ is $\Delta E = E_{\sigma^+} - E_{\sigma^-} = g\mu_B B = -2(\mu_l + \mu_v - \mu_c^b)B$, where $E_{\sigma^-}$ ($E_{\sigma^+}$) is the transition energies under $\sigma^-$ ($\sigma^+$) detection, $g$ is the exciton valley Landé g-factor, $\mu_l = 2\mu_B$ is the intra-atomic *d* orbital magnetic moment of valence band with Bohr magneton $\mu_B = 0.058$ meV/T and $\mu_v(\mu_c^b) = \frac{m_0}{m_h^*(m_e^*)}\mu_B$ denotes the valley orbital magnetic moments of valence (bottom conduction) band with $m_0$ the free-electron mass and $m_h^*$ ($m_e^*$) the effective mass of hole (electron)[41,44]. Evidently, the contributions of valley orbital magnetic moments from conduction and valence band compensate each other and the Landé g-factor of intralayer A excitons is negative[41-44]. Note that the intra-atomic orbital magnetic moment of the conduction band is zero. For spin-singlet and spin-triplet IXs, the valley Zeeman splitting is $\Delta E_{\text{spin-singlet}} = E_{\sigma^+} - E_{\sigma^-} = 2(\mu_l + \mu_v + \mu_c^t)B$ and $\Delta E_{\text{spin-triplet}} = E_{\sigma^+} - E_{\sigma^-} = -2(\mu_l + \mu_s + \mu_v + \mu_c^t)B$, respectively, where $\mu_s = 2\mu_B$ is the contribution from the spin magnetic moment and $\mu_c^t$ is the valley orbital magnetic moments of the top conduction band. It is clear that the Landé g-factor of spin-triplet IXs has the same sign with that of the intralayer A excitons, while the sign of Landé g-factor of spin-singlet IXs is opposite to that of the intralayer A excitons. The opposite sign of Landé g-factor between spin-singlet and spin-triplet IXs allows us to identify them unequivocally.

Figure 2B shows the valley Zeeman splitting of intralayer A excitons, IX2 (2.33 eV excitation) and IX4 (1.96 eV excitation) as a function of the applied magnetic field $B$ at 2K. Applying a linear fit, we extract the Landé g-factor of $g_A = -2 \pm 0.4$, $g_{IX2} = 5.9 \pm 0.3$ and $g_{IX4} = 6.2 \pm 0.1$. Remarkably, IX2 under 2.33 eV excitation and IX4 under 1.96 eV excitation exhibit almost the same Landé g-factor, including both the sign and magnitude. In addition, the sign of Landé g-factors of both IX2 and IX4 is opposite to that of intralayer A excitons. This strongly indicates that IXs under both 2.33 eV (IX1/IX2) and 1.96 eV excitations (IX3/IX4) are spin-singlet species[20,21,30]. It is worth stressing that the Landé g-factor of intralayer A excitons of bilayer MoS$_2$ (i.e., $g_A = -2 \pm 0.4$) contrasts the generally observed value of ~ −4 for other TMDCs (such as MoSe$_2$, WS$_2$ and WSe$_2$) described under the non-interacting independent-particle picture (that is, $m_h^* = m_e^*$ and $\mu_v = \mu_c$)[3,49-51]. This strongly indicates that the strong electron-electron exchange interactions are at play in bilayer MoS$_2$, driving the electron-hole symmetry breaking and leading to a net intercellular contribution from valley orbital magnetic moments[3]. Note that such unusual Landé g-factor of ~ −2 has also been uncovered previously in $h$-BN encapsulated monolayer and bilayer MoS$_2$[30,52]. In addition, if the bottom and top conduction bands share the same valley orbital magnetic moment, we can extract the valley orbital magnetic moments of valence ($\mu_v$) and conduction bands ($\mu_c$) in bilayer MoS$_2$ by combining the valley Zeeman splitting of intralayer A excitons and spin-singlet IXs: $\mu_c = \mu_c^t = \mu_c^b = \mu_B$ and $\mu_v = 0$. From a physical point of view, the zero $\mu_v$ is definitely wrong. Consequently, the valley orbital magnetic moments of bottom and top conduction bands should not be equal to each other, in good harmony with recent studies[53].

Is it possible that the IXs with higher energies under 2.33 eV excitation belong to a neutral type, while the IXs with lower energies under 1.96 eV excitation are charged species? Figure 2C and fig. S4 show the color map of PL spectra as a function of doping density in the absence of vertical electric field, excited by 1.96 eV and 2.33 eV continuous-wave lasers, respectively. Remarkably, IXs under both 2.33 eV (IX1/IX2) and 1.96 eV excitations (IX3/IX4) follow the similar doping-driven intensity evolution of neutral intralayer A excitons, demonstrating the neutral nature for both IX1/IX2 and IX3/IX4. In addition, the energy spacing of 40 meV between IX2 and IX4 (Fig. 1F), which is larger than that between intralayer A excitons and trions/Fermi polarons (26 meV), also excludes this hypothesis that IXs under 1.96 eV excitation are charged species, since the binding energies of IXs are smaller than that of intralayer excitons[24,28]. Considering that the energy spacing of 40 meV between IX2 and IX4 can match well the energy of phonon modes in $MoS_2$[54,55], one plausible origin for IX3/IX4 under 1.96 eV on-resonance excitation is phonon replicas. In addition, because of the inevitable existence of native defects, excitons bound to the intrinsic defect states may also be responsible for IXs under 1.96 eV excitation[56]. We stress that further experimental and theoretical studies are required to fully understand the detailed microscopic mechanism of the new IX species under 1.96 eV on-resonance excitation.

Figure 3A presents the circular polarization-resolved PL mapping of bilayer $MoS_2$ at 10 K under $\sigma^+$ (left panel) and $\sigma^-$ detections (right panel), excited by on-resonance $\sigma^+$ light of 1.96 eV. As evidenced by the strong contrast between the PL emission intensities under co- and cross-circularly polarized detections, IXs of bilayer $MoS_2$ can harbor robust spin-valley properties. Figure 3B shows the color plot of the degree of circular polarization, quantified as $\rho = \frac{I(\sigma^+) - I(\sigma^-)}{I(\sigma^+) + I(\sigma^-)}$, against photon energy and out-of-plane electric field, where $I(\sigma^\pm)$ is the measured intensity of left- (right-) handed circular-polarization component under $\sigma^+$ excitation of 1.96 eV. It is clearly shown that IXs possess robust negative circular polarization, whereas the intralayer A excitons have positive circular dichroism, indicating the opposite chirality between them. It is worth noting that due to the spin-layer locking effects and inversion symmetry breaking of individual local sectors, the non-vanishing circular polarization of intralayer A excitons is an intrinsic property of bilayer MoS₂, rather than caused by the extrinsic sample irregularities as claimed in the initial work[35,45,57,58]. More significantly, the magnitude of circular polarization of IXs is much larger than that of the intralayer A transitions. It is worth stressing that owing to the strong background from the intralayer A excitons under co-circularly polarized detection, the circular polarization of IXs shown in Fig. 3B is largely underestimated. Figure 3C presents the helicity-resolved PL spectra of bilayer MoS₂ at $E_z = \pm 0.103 MV/cm$. The IXs feature a strong signal under the cross-circularly polarized detection, but basically cannot be observed under the co-circularly polarized detection (also see in Supplementary Section 8), suggesting that the intrinsic degree of circular polarization can almost reach the theoretical limit (100%) for IXs under 1.96 eV on-resonance excitation.

The highly unusual close-to-unity, negative circular polarization of IXs in bilayer MoS₂ can be understood as the spin-preserving scattering of electron during the exciton formation and strongly suppressed valley depolarization. Upon the 1.96 eV excitation of σ⁺ circularly polarized photon, electron-hole pairs are generated in K valley of upper layer and -K valley of lower layer through spin-conserved interband intralayer A exciton transitions, as sketched by the green arrows in Fig. 3D. It is noteworthy that electron-hole pairs generated via IX transitions can be ignored because IXs with optical absorption belong to the species measured under the 2.33 eV excitation, and their corresponding energy is larger than 1.96 eV[24,28,30]. The imbalanced distribution of photo-generated carriers between layers within the same valley and between valleys within the same layer would cause relaxation. For instance, the photo-generated electrons in the lower conduction band of -K (K) valley of the lower (upper) layer would relax into the upper conduction band of K valley of the lower layer through valley phonons-assisted intervalley scattering (interlayer tunneling), while the relaxation of photo-generated holes is forbidden (Fig. 3E), considering that these two relaxation processes facilitate the spin-conserved scattering[59-61]. Such spin-preserving scattering of electrons would lead to the formation of σ⁻ polarized IXs (Fig. 3E), resulting in the observed negative circular polarization.

Based on recent studies[4,8,9], the strong electron-hole exchange interactions dominate the exciton valley depolarization in TMDCs through Maialle-Silva-Sham mechanism. For TMDCs in the strong scattering limit, the valley depolarization rate can be expressed as $\sim \bar{\Omega}^2 \tau$, where $\bar{\Omega}$ is the magnitude of the ensemble-averaged exchange interaction and proportional to the transition dipole, and $\tau$ is momentum relaxation time[4,7]. Owing to the spatially displaced wave functions of electrons and holes, the transition dipole and thus $\bar{\Omega}$ for IXs should be several times of magnitude smaller than that for intralayer excitons. As a consequence, Maialle-Silva-Sham valley depolarization mechanism is largely suppressed for IXs, resulting in the long valley lifetime and giant circular polarization.

In addition, we note that another new IX species with energy above 15 meV of IX4 but weak intensity, dubbed as IX5, can be observed under cross-circularly polarized detection at certain positive electric fields, excited by 1.96 eV radiation (Fig. 3A). Being akin to IX4, IX5 also features negative circular polarization. Considering that the energy difference between IX5 and IX2 (25 meV) is comparable to that between intralayer A excitons or trions (26 meV), IX5 may belong to a charged IX and deserves further studies.

Apart from the optically addressable valley polarization, the valley degree of freedom of IXs can also be controlled by breaking the time-reversal symmetry with an out-of-plane magnetic field[3,41]. Importantly, the enhanced Landé *g*-factor and valley Zeeman splitting of IXs compared to intralayer A exciton in bilayer $MoS_2$ indicate a more effective manipulation of the valley polarization through magnetic fields. Figure 4A shows the helicity-resolved magneto-PL spectra under cross-circularly polarized detections at 0T, ±3T and ±8.5T for device D3, excited by 1.96 eV laser light. Note that a small out of plane electric field is applied to drag IXs into our detection range. At 0T, $\sigma^+$ and $\sigma^-$ polarized components share the same intensity. When applying a magnetic field, the low-energy components exhibit stronger intensity than the high-energy components, indicating the large formation rate of low-energy IXs and the spontaneous circular polarization (Supplementary Section 11). Figure 4B shows the spontaneous circular polarization under different magnetic fields, defined by $\rho = \frac{I(\sigma^+) - I(\sigma^-)}{I(\sigma^+) + I(\sigma^-)}$. The spontaneous valley polarization increases linearly with the increase of the magnetic fields: $\rho = \alpha B$, where $\alpha$ indicates the magnetic modulation factor. Through fitting, we extract $\alpha = \sim 7.7\% \cdot T^{-1}$, which, to the best of our knowledge, is the largest value among TMDCs and their heterostructures[15,48,49,52,62-64] (Fig. 4C). Such effective manipulation of valley degree of freedom by magnetic field in bilayer $MoS_2$ would pave the way for transformative valleytronic devices.

To summarize, we observe two new IX species with large out-of-plane electric dipole in bilayer $MoS_2$, under 1.96 eV on-resonance excitation. The valley degree of freedom of such new IXs in bilayer $MoS_2$ can be manipulated through both circularly polarized light and magnetic field. Because of the combination of spin-preserving scattering of an electron during the formation and strongly suppressed valley depolarization, IXs in bilayer MoS$_2$ feature a close-to-unity, but negative circular polarization. Moreover, by breaking the valley degeneracy with an out-of-plane magnetic field, we identify a record level of spontaneous valley polarization for IXs in bilayer MoS$_2$. Such highly tunable IXs in bilayer MoS$_2$ with optically and magnetically addressable valley pseudospin hold great promise for the engineering of valley-functional chiral electronic/optoelectronic devices and quantum-computation.

## Methods

### Sample fabrication

We prepare *h*-BN encapsulated bilayer MoS$_2$ with few-layer graphene (FLG) as bottom and top gate by van der Waals-mediated dry transfer approach employing propylene carbonate (PC) stamp. First, FLG, bilayer MoS$_2$ and *h*-BN flake are exfoliated onto SiO$_2$ (300 nm)/Si substrate. And then PC stamps are used to pick up *h*-BN flake, bilayer MoS$_2$ and another *h*-BN flake in sequence with accurate alignment based on an optical microscope. The sandwiched *h*-BN/bilayer MoS$_2$/*h*-BN heterostructure is then transferred onto the FLG which serves as a bottom gate, and another FLG is transferred on it as a top gate by dry transfer technique. Finally, metal contacts to the bilayer MoS$_2$ and FLG are patterned by the standard micro-fabrication processes including e-beam lithography (EBL), metal evaporation Ti (3 nm)/Au (40 nm) and lifting-off.

### Photoluminescence experiment

In our experiments, devices are wire-bonded onto a chip carrier and placed in a homemade optical chamber. The sample temperature, unless otherwise specified, is T=10K. PL spectra are obtained using a HORIBA spectrometer (LabRAM HR Evolution) in a confocal backscattering configuration. Light from 633 nm (1.96 eV) and 532 nm (2.33 eV) continuous laser is focused through a Nikon objective (N.A.=0.5 W.D.=10.6 F.N.=26.5) onto the sample with a spot diameter of ~1.5 μm. For helicity-resolved PL measurements, the excitation laser is first guided through a vertical linear polarizer followed by a quarter-wave plate to achieve σ+ circular polarization. The circular polarization of the excitation laser is confirmed at the sample position. The backscattered PL signals going through the same quarter-wave plate are collected and analyzed with a half-wave plate and a linear polarizer.

### Acknowledgements

We would like to thank Wang Yao in University of Hong Kong for fruitful discussions.

**Figures and figure captions**

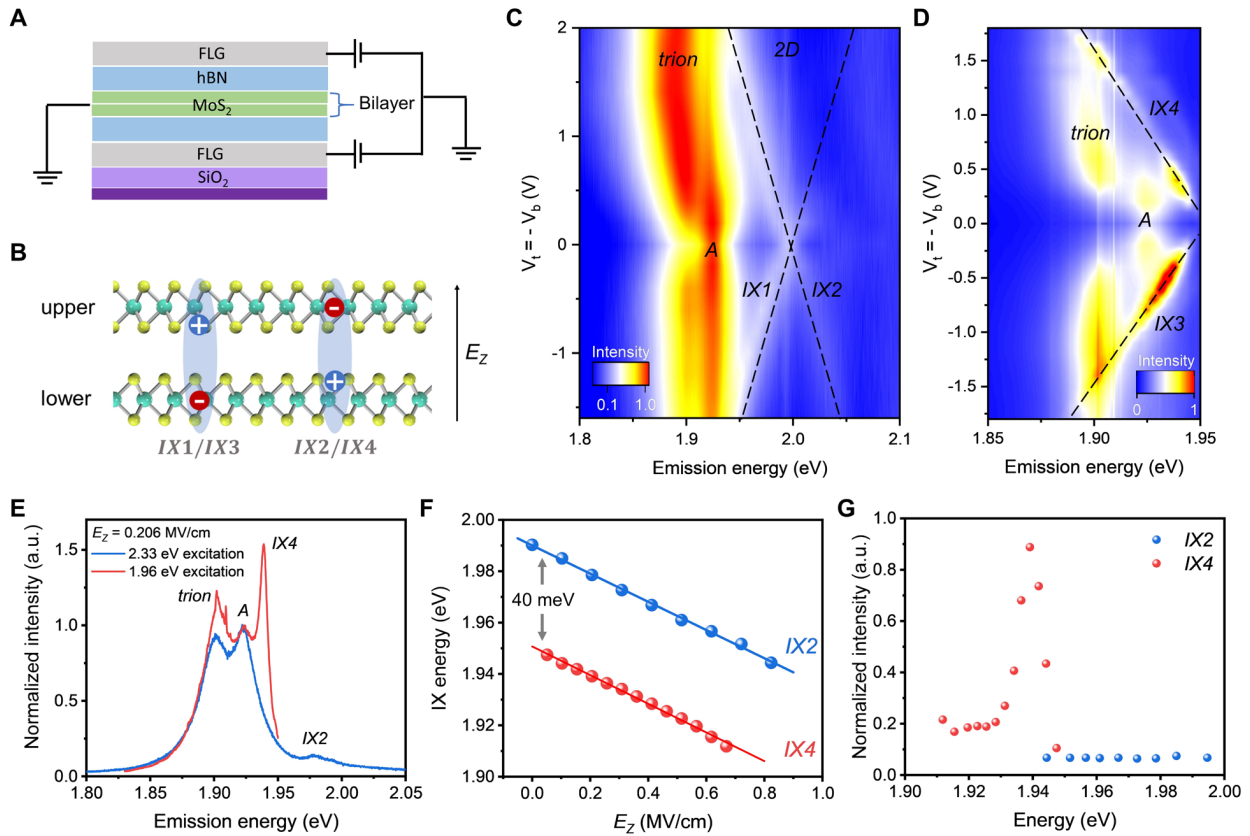

**Fig. 1. Observation of IXs in bilayer MoS₂.** (A) Schematic structure of *h*-BN encapsulated dual-gate bilayer MoS$_2$ device. (B) Schematic diagram of the two neutral IXs with electron localized in lower and upper layers in bilayer MoS$_2$. (C) Color map of PL spectra as a function of electric field under 2.33 eV excitation. (D) Color map of PL spectra as a function of electric field under 1.96 eV excitation. (E) Comparison of PL spectra at specific electric field under 2.33 eV and 1.96 eV excitation, the PL intensity of A$^0$ is normalized. (F) Energies of IX2 and IX4 as a function of positive out-of-plane electric field. (G) Laser power normalized PL intensity of IX2 and IX4 as a function of energy.

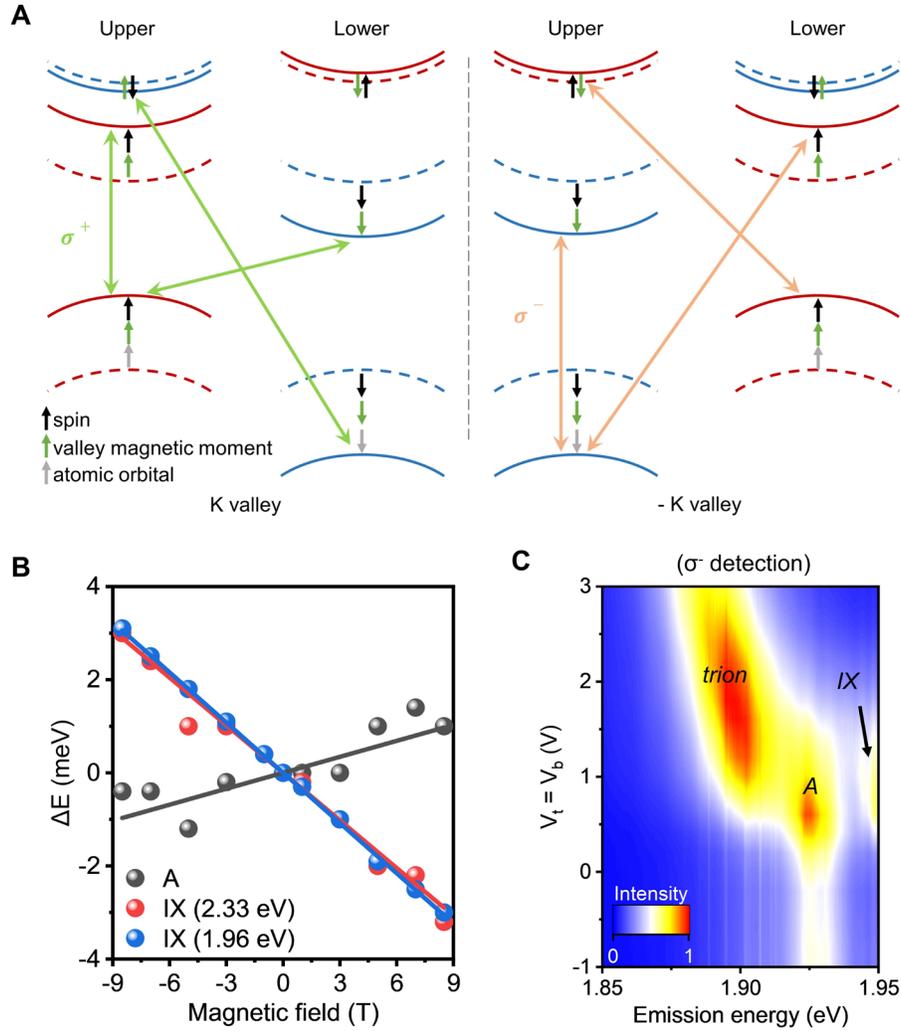

**Fig. 2. Neutral spin-singlet IX in bilayer MoS$_2$ under 1.96 eV excitation.** (A) Energy level diagram showing the three contributions to the valley Zeeman shifts of intralayer A exciton, spin-singlet and -triplet IXs in bilayer MoS$_2$. The corresponding optical selection rules are also shown, for the sake of clarity in K (-K) valley only right-handed (left-handed) helicity is illustrated. (B) Valley Zeeman splitting of neutral IXs and intralayer exciton under magnetic field. (C) Color map of PL spectra as a function of doping density under right-handed ($\sigma^+$) excitation at 1.96 eV, the IX shares the similar evolution with neutral intralayer A exciton.

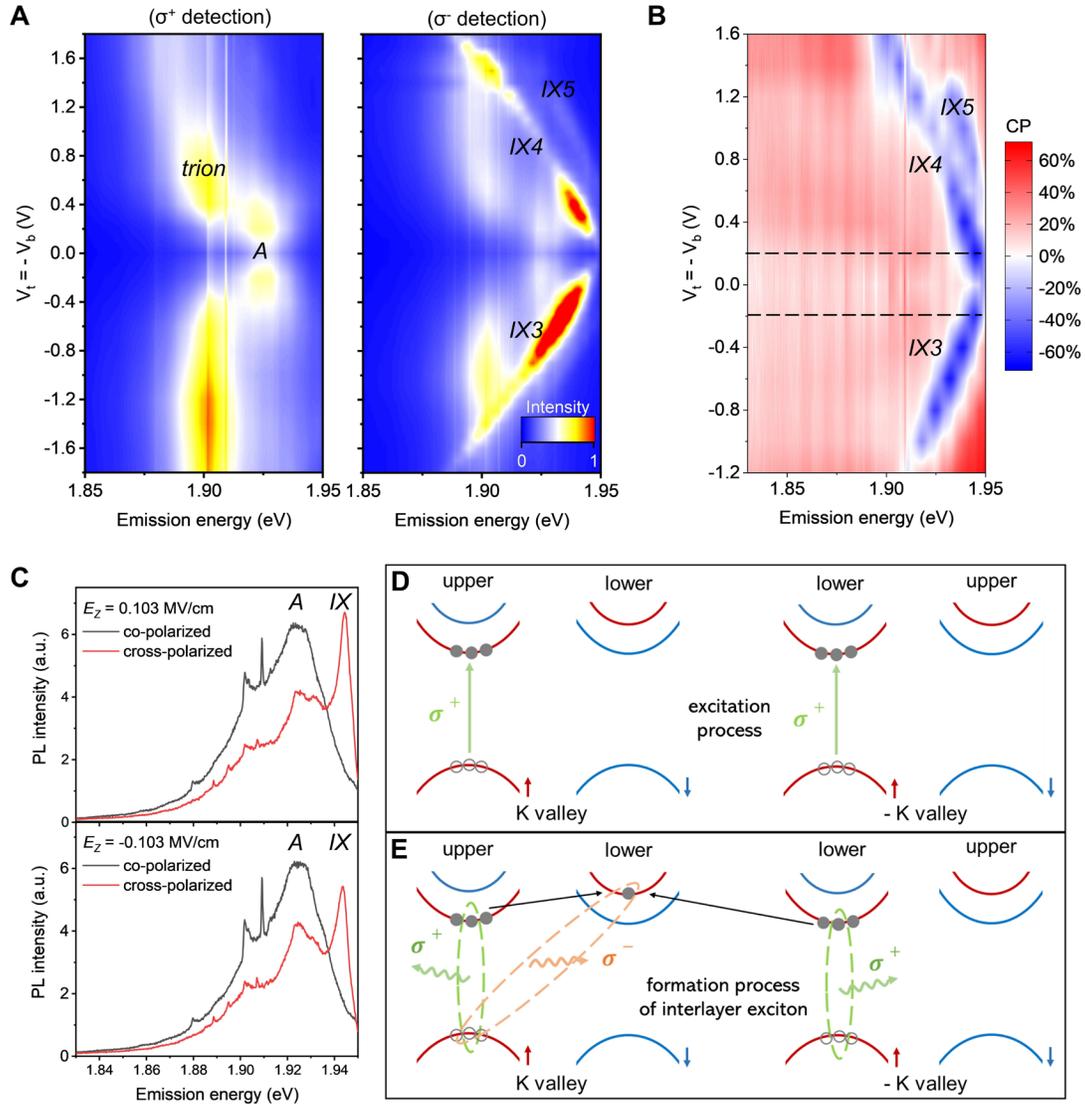

**Fig. 3. Unity negative circular polarization of IXs in bilayer MoS$_2$ under 1.96 eV excitation.** (A) Color map of PL spectra as a function of electric field under co- (left panel) and cross-polarization (right panel) detection. The excitation light is right-handed ($\sigma^+$) at 1.96 eV. (B) Color map of circular polarization as a function of the out-of-plane electric field under 1.96 eV excitation. (C) Single PL spectra under electric field as the dot black line shown in Fig. 3B. The intensity of IXs is basically negligible under co-circularly polarized detection, which indicates a close-to-unity negative circular polarization. (D) Helicity-dependent excitation process. (E) IXs formation process and helicity-dependent recombination process.

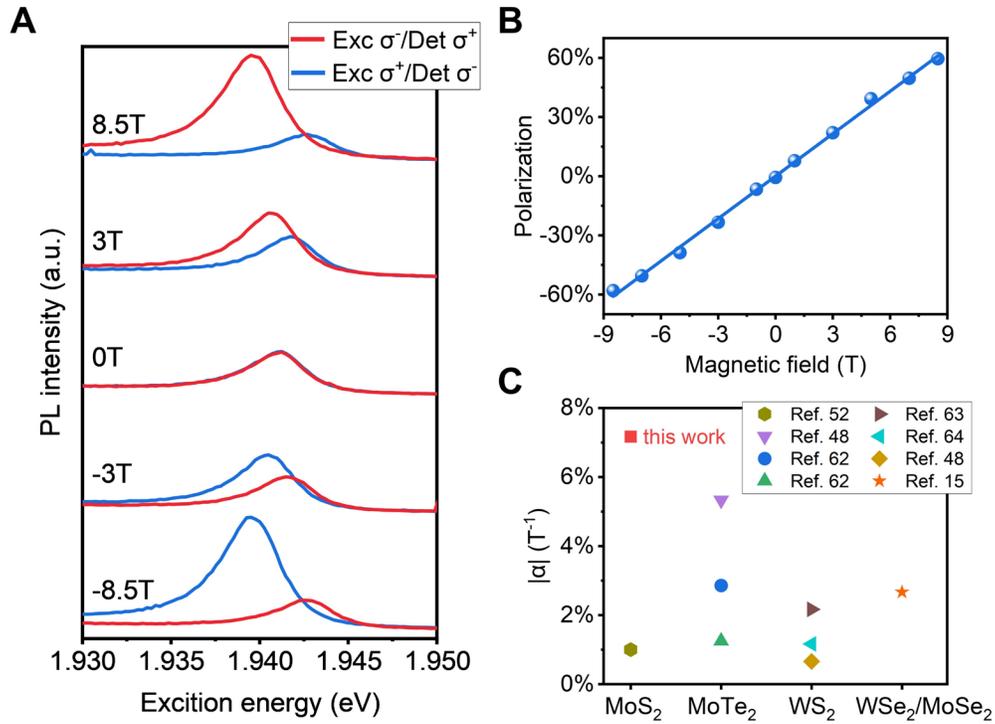

**Fig. 4. Valley polarization of IXs in bilayer MoS$_2$ under magnetic field.** (A) Single PL spectra at specific magnetic field under cross-circularly polarized detection, the excitation energy is 1.96 eV. (B) Spontaneous valley polarization of neutral IXs and intralayer exciton under magnetic field. (C) Comparation of magnetic field modulated spontaneous valley polarization between IX3 in pristine bilayer MoS$_2$ we detected with excitons in TMDCs and their heterostructures in literature.

# Supplementary Materials for

# Optically and magnetically addressable valley pseudospin of interlayer excitons in bilayer MoS$_2$

Yanchong Zhao[1,2], Luojun Du[3,*], Shiqi Yang[4,5], Jinpeng Tian[1,2], Xiaomei Li[1,2], Cheng Shen[1,2], Jian Tang[1,2], Yanbang Chu[1,2], Kenji Watanabe[6], Takashi Taniguchi[7], Rong Yang[1,8,9], Dongxia Shi[1,2,8], Zhipei Sun[3,10], Yu Ye[4,*], Wei Yang[1,2,9,*], Guangyu Zhang[1,2,8,9,*]

*Corresponding author. Email: luojun.du@aalto.fi; ye_yu@pku.edu.cn; wei.yang@iphy.ac.cn; gyzhang@iphy.ac.cn

## Section S1: Device information

Figure S1 shows one of the typical dual-gate h-BN encapsulated bilayer MoS$_2$ devices (D1) in our experiment. In this device, there are two different bilayer MoS$_2$ flakes, and they can be modulated individually, giving similar behavior. The thickness of top and bottom h-BN is 9.47 nm and 9.49 nm, respectively.

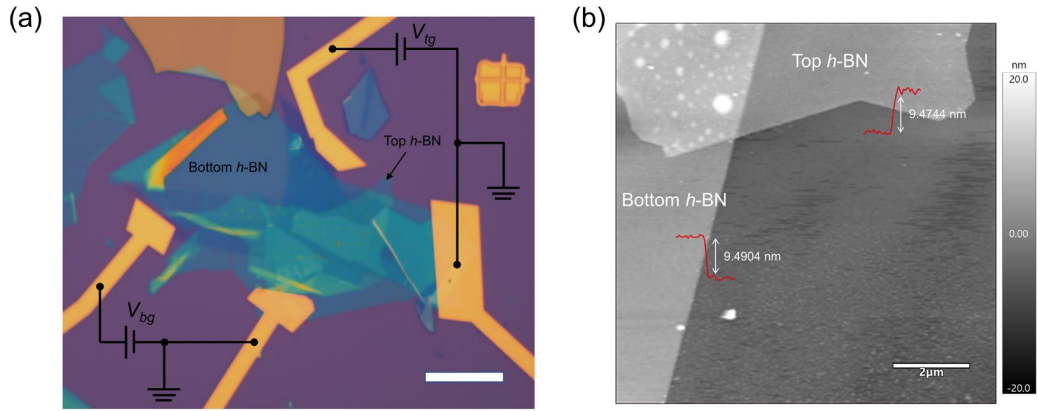

**Fig. S1. a**, Optical image of a dual-gate h-BN encapsulated bilayer MoS$_2$ device. Scale bar, 20 μm. **b**, The corresponding atomic force microscopy (AFM) image of the device in **a**, showing the thickness of top and bottom h-BN flake is 9.47 nm and 9.49 nm, respectively.

## Section S2: Interlayer hole hybridization in 2H-stacked TMDC homobilayers

Due to the non-vanishing interlayer hopping $2t_\perp$ (0.086 eV) for holes, electronic states with the same spin configuration between the valence bands of upper and lower layers in bilayer TMDC are hybridized. The interlayer hole hybridization can be described by a two-level Hamiltonian model: $\begin{bmatrix} \lambda & t_\perp \\ t_\perp & -\lambda \end{bmatrix}$, and gives rise to two hybridized eigenstates with energy separation of $2\sqrt{\lambda^2 + t_\perp^2}$: $|\psi_+\rangle = \frac{1}{\sqrt{2}}(\cos\alpha\,|\psi_u\rangle + \sin\alpha\,|\psi_l\rangle)$ and $|\psi_-\rangle = \frac{1}{\sqrt{2}}(\sin\alpha\,|\psi_u\rangle - \cos\alpha\,|\psi_l\rangle)$, where basis $|\psi_u\rangle/|\psi_l\rangle$ is the wave-function of decoupled upper/lower layer, $2\lambda$ (0.147 eV) denotes the spin-orbit splitting in the valence band, and $\cos 2\alpha = \frac{\lambda}{\sqrt{\lambda^2 + t_\perp^2}}$ (Fig. S2a). The hybridized hole states in bilayer TMDC possess components from both the upper and lower layers and enable the formation of IXs with the contribution of intralayer bright excitons (Fig. S2b). Consequently, the IXs harbor strong oscillator strength and hold great promise for strongly coupled exciton-polaritons and optoelectronics. Based on the hybridized hole wavefunctions, we can know that the in-built electric dipole moment for IXs is $ed\cos\alpha = 0.605\ e\ nm$.

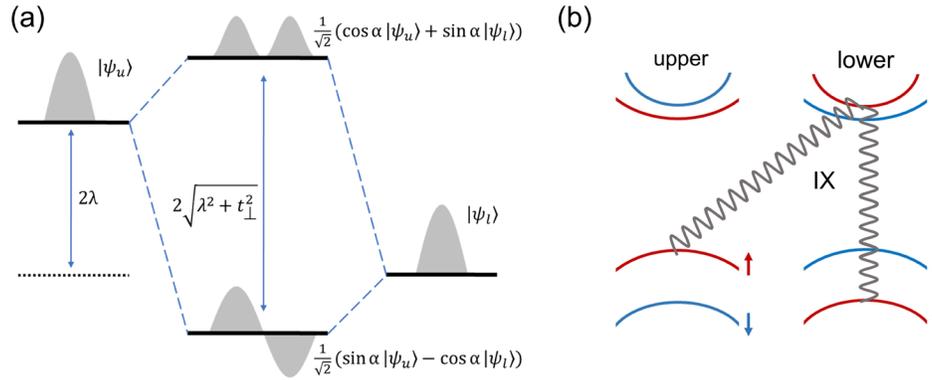

**Fig. S2. a**, Schematic diagram of the interlayer hole hybridization in bilayer TMDC. **b**, Illustration of spin-singlet IX, an electron localized on lower layer interacts with a hybridized hole state to form a spin-singlet IX. For clarity, only the spin-up IX is shown; there is also a spin-down IX with the same energy.

**Section S3: Temperature dependent emission energy under 2.33 eV excitation**

Figure S3a shows the temperature-dependent PL spectra of bilayer $MoS_2$ at zero gate voltage, excited by a linearly polarized 2.33 eV continuous-wave laser. We can see that, besides the well-studied neutral intralayer A exciton ($A^0$) transition around 1.92 eV, a prominent peak about 70 meV above $A^0$ exciton can be clearly identified, even at room temperature (inset of Fig. S3a). The transition around 2.0 eV is assigned to IX of bilayer $MoS_2$. Figure S3b shows the temperature-dependent emission energies of both $A^0$ (black dot) and IX (red dot), and the corresponding solid lines are fitted by the standard semiconductor bandgap dependence using $E_g(T) = E_g(0) - $

$S\hbar\omega[\coth(\frac{\hbar\omega}{2kT}) - 1]$, where $E_g(0)$ is the ground-state transition energy at 0 K, S, $\hbar\omega$ and $k$ are dimensionless coupling constant, average phonon energy and Boltzmann constant. From the fits, we extract for $A^0$ (IX) the $E_g(0)$ = 1.924 (1.994) eV, S = 1.89 (1.96) and $\hbar\omega$ = 21.33 meV for both.

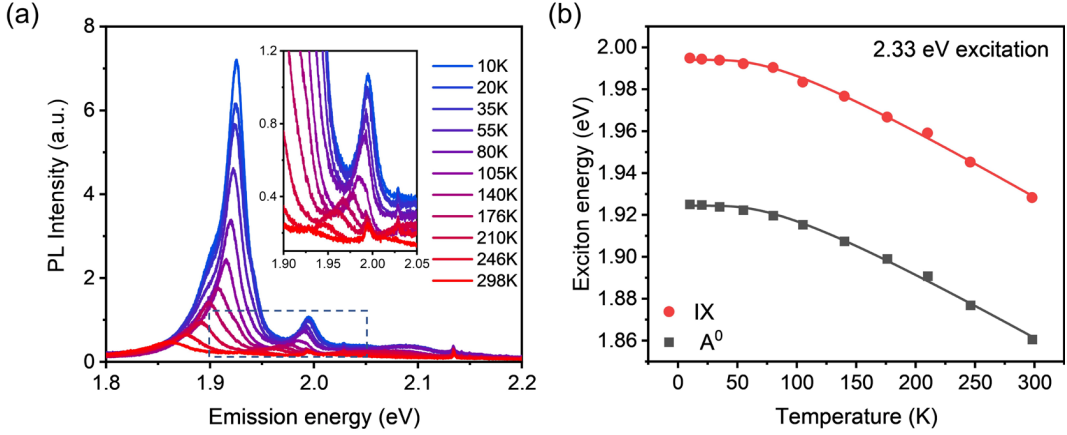

**Fig. S3. a**, Temperature dependent PL spectra under 2.33 eV excitation. Inset is the zoom-in of the blue dot area. **b**, Temperature dependent emission energies of intralayer $A^0$ exciton and IX.

**Section S4: Electrostatic model**

The dual-gate $h$-BN encapsulated bilayer MoS$_2$ device allows us to individually apply an out-of-plane electric field or carrier density (doping) to the bilayer MoS$_2$ using the voltages applied to the top and bottom gates.

The out-of-plane electric field is defined by

$$E_Z = \frac{C_{b-BN}V_{BG} - C_{t-BN}V_{TG}}{2\varepsilon_{biMoS_2}}$$

where

$$C_{b-BN} = \varepsilon_0\varepsilon_{hBN}/d_{b-BN}$$

and $C_{t-BN} = \varepsilon_0\varepsilon_{hBN}/d_{t-BN}$

are the geometric capacitance of the bottom and top $h$-BN, $\varepsilon_0$ is the vacuum permittivity and $\varepsilon_{hBN} \approx 3.5$ is the dielectric constant of $h$-BN. So, the out-of-plane electric field can be written as

$$E_Z = \frac{\varepsilon_{hBN}}{2\varepsilon_{biMoS_2}}(\frac{V_{BG}}{d_{b-BN}} - \frac{V_{TG}}{d_{t-BN}})$$

And from the plate capacitor model the total carrier density in bilayer MoS$_2$ is

$$n_{tot} = \frac{1}{e}\varepsilon_0\varepsilon_{hBN}(\frac{V_{BG}}{d_{b-BN}} + \frac{V_{TG}}{d_{t-BN}})$$

**Section S5: Doping concentration dependence of interlayer exciton under 2.33 eV excitation**

Figure S4 presents the PL spectra at different doping concentrations under 2.33 eV excitation. IX is strongly quenched under electron doping and shares a similar doping-driven intensity evolution with neutral intralayer A exciton ($A^0$), indicating that the observed IX should be a neutral one.

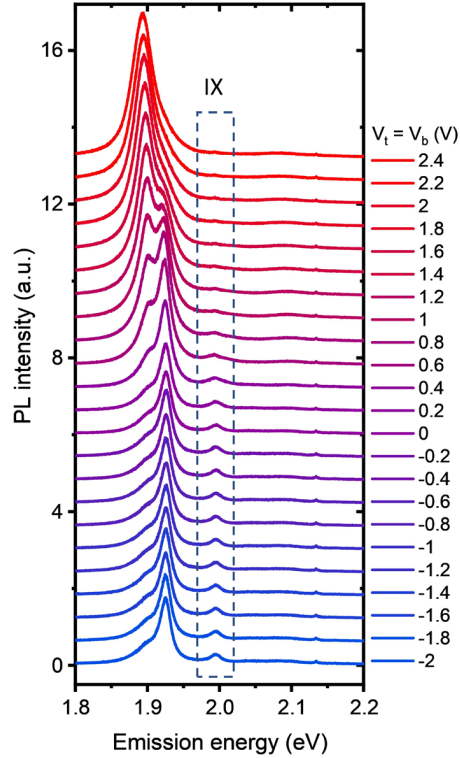

**Fig. S4** Doping dependent PL spectra of bilayer MoS$_2$ under 2.33 eV excitation at 10 K. The IX shares the similar evolution with neutral intralayer A exciton.

**Section S6: The polarization selection rules for the intralayer and interlayer transitions**

The interband transitions between the conduction band $\psi_{c,S_z'}$ and valence band $\psi_{v,S_z}$ are characterized by the optical matrix element $\langle \psi_{c,S_z'} | \hat{P}_\pm | \psi_{v,S_z} \rangle$, where $S_z$ is the spin index and $\hat{P}_\pm = \hat{P}_x \pm i \hat{P}_y$ couples to a $\sigma^+$ ($\sigma^-$) circularly polarized photon with $\hat{P}_x$ and $\hat{P}_y$ the $x$- and $y$-component of the momentum operator, respectively[1, 2]. Using $\langle \widehat{C_3} | \hat{P}_\pm | \widehat{C_3^{-1}} \rangle = e^{\mp \frac{i 2\pi}{3}} \hat{P}_\pm$ and $\langle \widehat{\sigma_h} | \hat{P}_\pm | \widehat{\sigma_h^{-1}} \rangle = \hat{P}_\pm$, we obtain:

$$\langle\psi_{c,S_z'}|\hat{P}_\pm|\psi_{v,S_z}\rangle = e^{\frac{i2\pi}{3}(C_3(c)-C_3(v)+S_z'-S_z\mp 1)}\langle\psi_{c,S_z'}|\hat{P}_\pm|\psi_{v,S_z}\rangle$$
$$= \sigma_h(m)\sigma_h(n)e^{i\pi(S_z'-S_z)}\langle\psi_{c,S_z'}|\hat{P}_\pm|\psi_{v,S_z}\rangle$$

where $\widehat{C_3}$ and $\widehat{\sigma_h}$ are the operators of three rotational symmetry and the out-of-plane mirror symmetry, respectively[1]. The above equation imply that the interband transitions satisfy: $e^{\frac{i2\pi}{3}(C_3(c)-C_3(v)+S_z'-S_z\mp 1)} = 3N$. And for intralayer A excaitons, $S_z' = S_z$. While the spin restriction is removed for the IXs due to the breaking of the out-of-plane mirror symmetry. Using the known $C_3(c), C_3(v), S_z'$ and $S_z$ in the previous literatures[3], we obtain the selection rules for the intralayer and interlayer transitions in bilayer MoS$_2$, as shown in Figure S5.

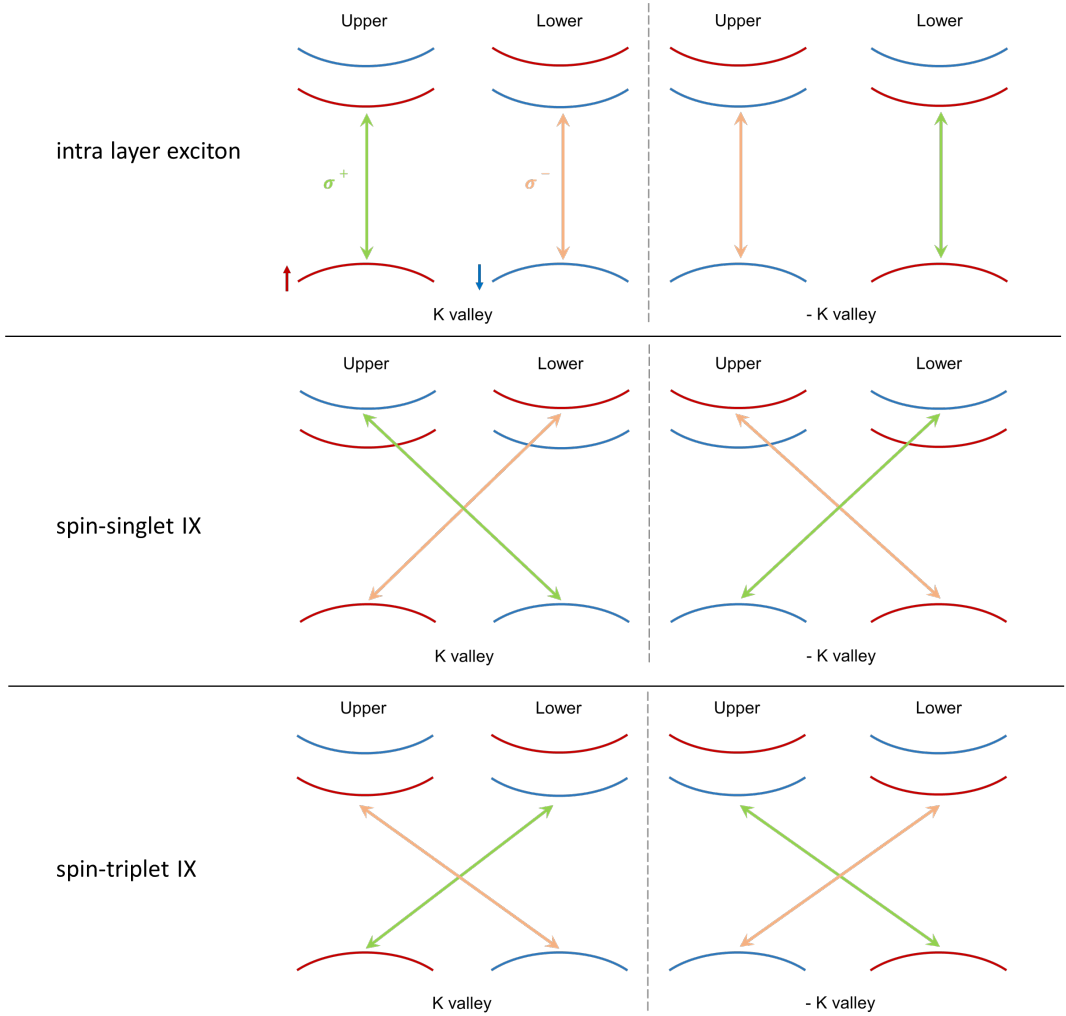

**Fig. S5.** polarization selection rules for the intralayer and interlayer transitions.

**Section S7: Circular polarization of interlayer exciton under 2.33 eV excitation**

Figure S6a shows the electric field-dependent co- (red solid line) and cross- (black solid line) circularly polarized emission signals in PL measurements under 2.33 eV excitation. For the IX, it shows a small circular polarization (~ 10%) under 2.33 eV excitation because of the large intervalley scatter. As shown in Fig. S6b we plot the detailed optical transition in the K and -K valley without (top) and with a finite electric field (bottom), and such property is consistent with our spin singlet exciton picture.

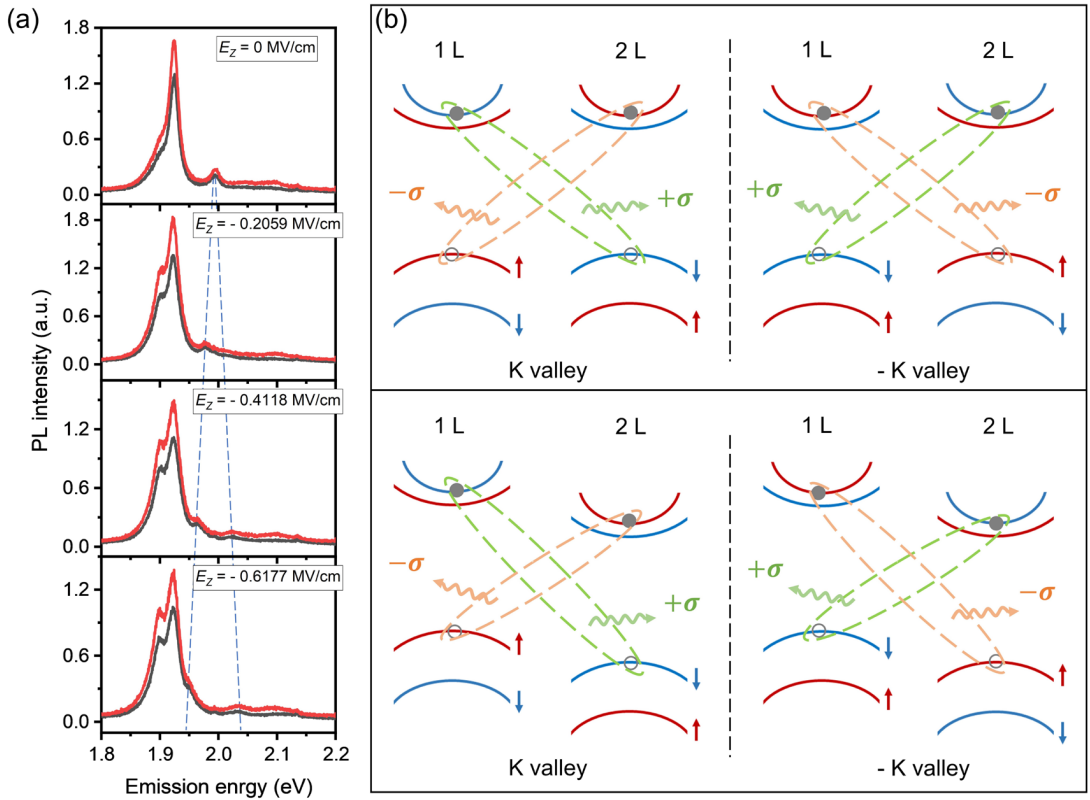

**Fig. S6. a**, Co- and cross-circular polarized detection of the spin-singlet emission in PL measurement under different out of plane electric field. **b**, Detailed optical transition of the spin-singlet under 2.33 eV excitation.

**Section S8: Circular polarization of interlayer exciton under 1.96 eV excitation**

Under the on-resonance condition of 1.96 eV excitation, we study the polarization-resolved PL spectra at different doping densities. Interlayer exciton (IX) emerges as a new exciton emission peak near 1.95 eV. The doping-dependent co- (Fig S7a) and cross- (Fig S7b) circular polarized detection results show that the emission only exists under the cross-circular polarized channel, indicating a large negative circular polarization of this interlayer exciton. Besides, IX also shares the same doping dependence as the neutral intralayer A excitons.

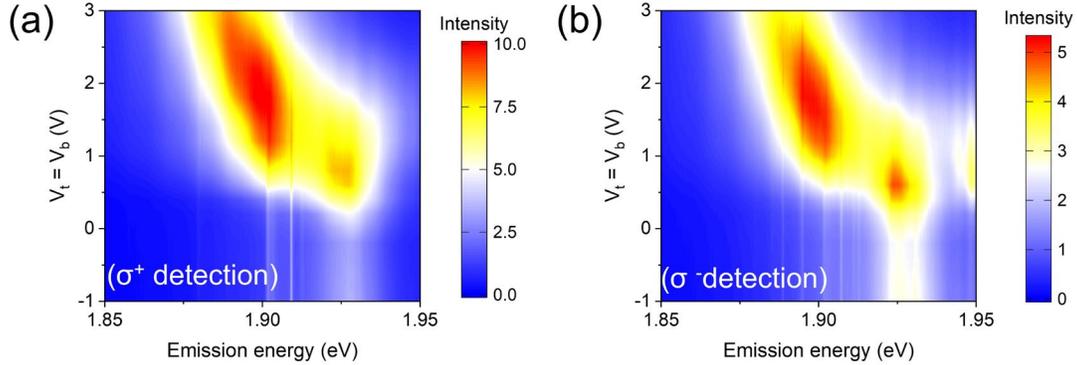

**Fig. S7.** Doping density dependent PL spectra under co- and cross- circular polarization detection, excited by σ⁺ circularly polarized radiation with energy of 1.96 eV.

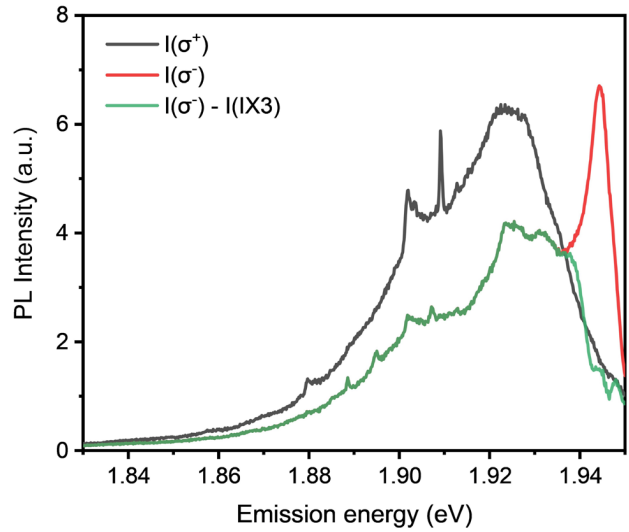

**Fig. S8.** Close-to-unity circular polarization.

Figure S8 presents the polarization-resolved PL spectra of bilayer MoS$_2$ under an electric field of $E_z = 0.103 MV/cm$, excited by 1.96 eV σ⁺ circularly polarized radiation. The co- (black solid line) and cross- (red solid line) circular polarized detection results indicate that the emission from IX3 only exists under the cross-circular polarized channel. We then extract the IX3 by fitting and deduct it from the cross-circular polarized detection PL spectra, and the result (green solid line) shows nearly equal intensity with the co-circular polarized detection near the energy of spin-triplet IX. It demonstrates that the intensity of IX3 is basically negligible under co-circularly polarized detection, suggesting the IX3 can harbor a negative close-to-unity circular polarization.

**Section S9: Magneto photoluminescence under 2.33 eV**

Figure S9a shows polarized PL spectra of bilayer $MoS_2$ under 2.33 eV excitation, at specific magnetic field IX shows opposite Zeeman shift compared with intralayer A exciton suggesting its spin-singlet property (Fig. S9b). Figure S9c demonstrates the valley Zeeman shift of conduction and valence band induced by the magnetic field, it shows that bare spin, valley magnetic moment, and atomic orbital all contribute to the shift of band edge. The dashed (solid) lines are the conduction and valence band edges at zero (positive) magnetic field, with blue and red denoting spin up and down, respectively.

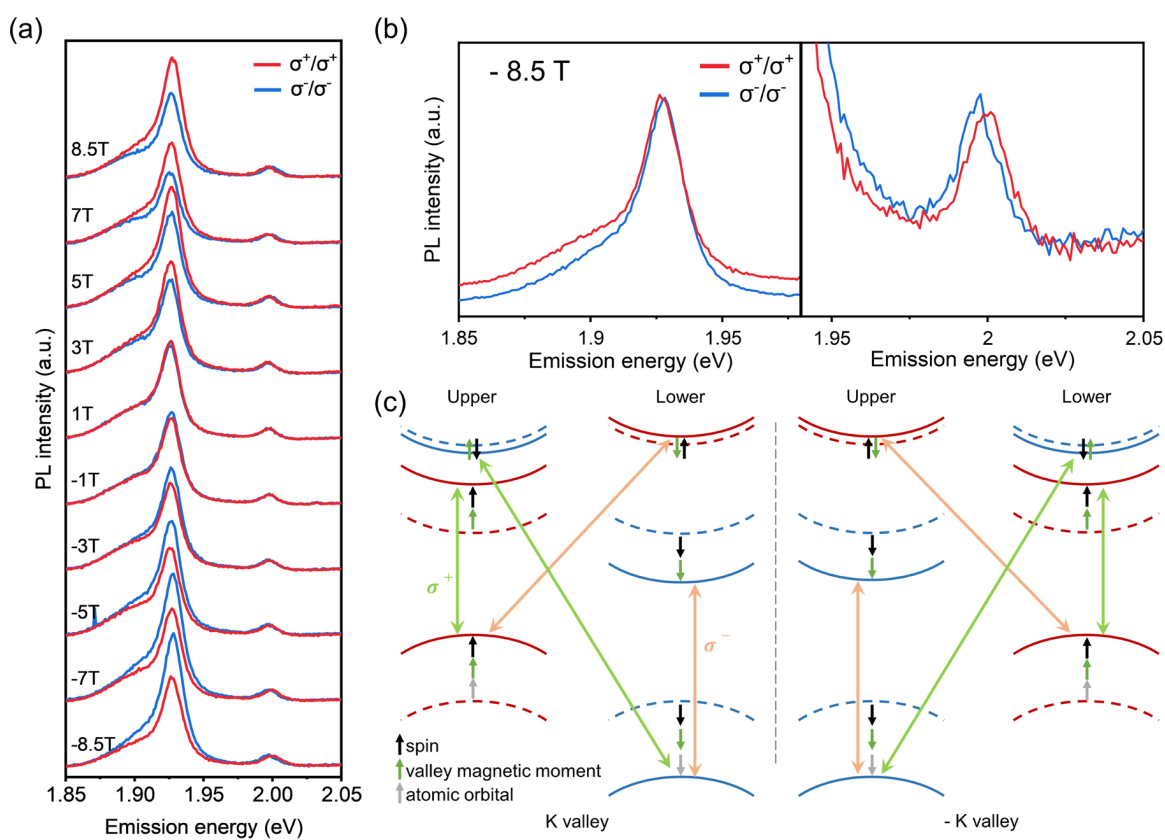

**Fig. S9. a**, Single PL spectra at specific magnetic field under co-circularly polarized detection, the excitation energy is 2.33 eV. **b**, Polarization-resolved valley-exciton PL of intralayer A exciton (left) and spin-singlet IX (right) at – 8.5 T. The valley Zeeman shift of intralayer A exciton under magnetic field is not obvious because of small Landé g factor, the intensity of intralayer A exciton is normalized. **c**, Energy level diagram showing the three contributions to the valley Zeeman shifts of intralayer A exciton and spin-singlet IX in bilayer $MoS_2$ (black for spin, green for valley magnetic moment, gray for atomic orbital).

## Section S10: Excitation power dependence of IX3

The interlayer exciton IX3 emission exhibits distinct power dependence compared to the known neutral intralayer A exciton. Figure S10 presents the normalized PL spectra under 1.96 eV with powers ranging from 5 μW to 2000 μW. The IX3 emission dominates the spectrum at low powers, but gradually saturates as the power increases. The strong power saturation does not conform to the hallmark of biexcitons[4-7]. Note that in order to drag IX3 into our detection range, a small out-of-plane electric field is applied.

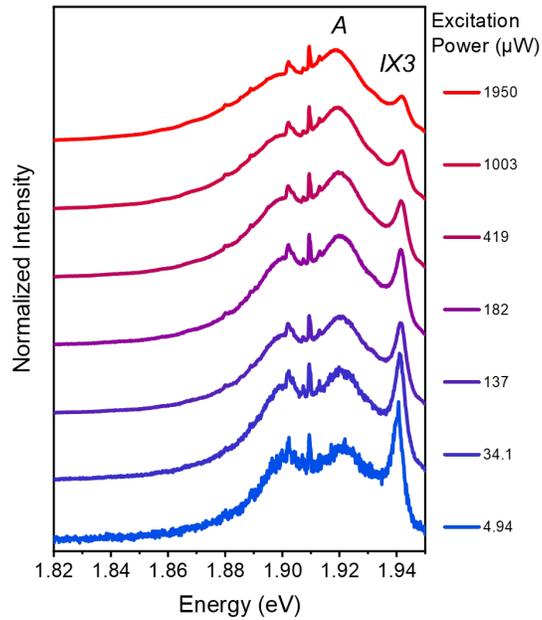

**Fig. S10.** Normalized PL spectra of IX3 at selected excitation powers.

## Section S11: Valley exciton energy in a magnetic field

Our results demonstrate that for IXs the lower-energy exciton retains more valley polarization than the higher-energy exciton under magnetic field, such behavior is contrary to the neutral intralayer excitons in TMDC monolayer that have been studied widely. The opposite valley spontaneous polarization phenomenon under magnetic field between intralayer A exciton and IXs in bilayer $MoS_2$ can be understood as their different exciton dispersions. For the intralayer A exciton (Fig. S11a), exchange interaction between electrons and holes induced the exciton dispersion splitting into two branches[8,9], and the upper branch has a steeper dispersion. At $B = 0$, the two branches touch at the energy minimum, a finite magnetic field will cause valley Zeeman to shift and lift this degeneracy. After excitation, the steeper dispersion of the upper exciton branch facilitates the formation of excitons (thickness of the dotted line) because smaller momentum transfers are required, resulting in the larger intensity of the higher energy exciton state[8, 10]. On the contrary, the exchange interaction is rather weak for IXs (Fig. S11b). Consequently, the two branches have almost the same dispersion and are totally degenerate at $B = 0$. After that, a finite magnetic field will also lift the two branches, and more excitons are formatted in the lower energy exciton state, resulting in the intensity of the lower energy exciton state.

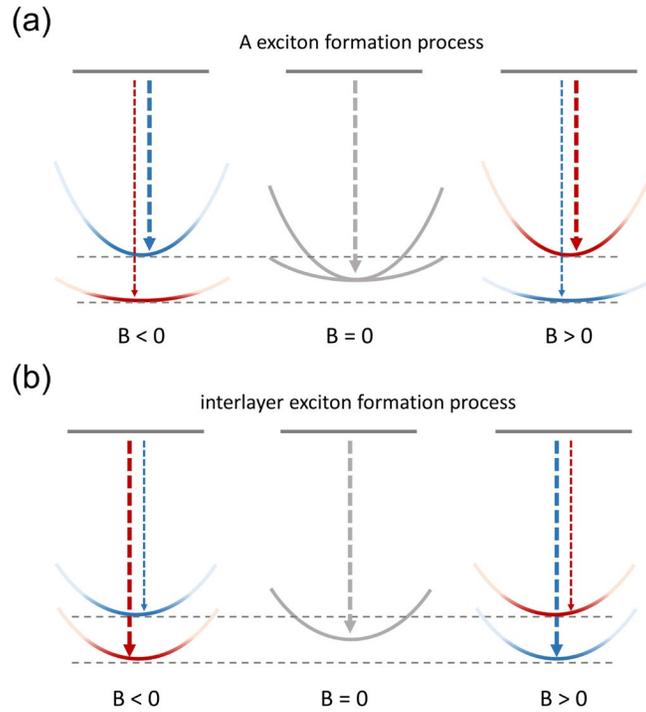

**Fig. S11.** Dispersions of exciton energy spectrum with and without a magnetic field of intralayer A exciton (**a**) and interlayer exciton (**b**). Gray curves represent the superposition of σ+ and σ−, and blue (red) denotes σ+ (σ−). Dotted line represents for the exciton formation after excitation, the thickness of dotted line indicates the formation rate to the exciton ground state before recombination.